# Reflections on the spatial performance of atom probe tomography in the analysis of atomic neighbourhoods

Baptiste Gault[a,b], Benjamin Klaes[c], Felipe F. Morgado[a], Christoph Freysoldt[a], Yue Li[a], Frederic De Geuser[d], Leigh T. Stephenson[a], François Vurpillot[c]

[a] Max-Planck-Institut für Eisenforschung, Max-Planck-Str. 1, 40237 Düsseldorf, Germany.
[b] Department of Materials, Royal School of Mines, Imperial College, Prince Consort Road, London SW7 2BP, United Kingdom.
[c] Groupe Physique des Matériaux, Université de Rouen, Saint Etienne du Rouvray, Normandie 76800, France
[d] Univ. Grenoble Alpes, CNRS, Grenoble INP, SIMAP, 38000 Grenoble, France

## Abstract

Atom probe tomography is often introduced as providing 'atomic-scale' mapping of the composition of materials and as such is often exploited to analyse atomic neighbourhoods within a material. Yet quantifying the actual spatial performance of the technique in a general case remains challenging, as they depend on the material system being investigated as well as on the specimen's geometry. Here, by using comparisons with field-ion microscopy experiments and field-ion imaging and field evaporation simulations, we provide the basis for a critical reflection on the spatial performance of atom probe tomography in the analysis of pure metals, low alloyed systems and concentrated solid solutions (i.e. akin to high-entropy alloys). The spatial resolution imposes strong limitations on the possible interpretation of measured atomic neighbourhoods, and directional neighbourhood analyses restricted to the depth are expected to be more robust. We hope this work gets the community to reflect on its practices, in the same way, it got us to reflect on our work.

## 1. Introduction

Atom probe tomography (APT) stems from the field-ion microscopy (FIM) that allowed Erwin Müller and his collaborators to image individual atoms already in the 1950s (Müller & Bahadur, 1956). In FIM, the image of each of the surface atoms is formed by the successive impact of thousands of gas ions per second on the screen. The ion trajectories in FIM are expected to be determined, for a given specimen geometry and microscope, only by the distribution of the electrostatic field (Smith & Walls, 1978), assuming that dynamic effects associated with the pulsed voltage can be neglected. To a first approximation, the projection of these ions can be well reproduced by an equidistant projection (Wilkes et al., 1974), or, to a certain extent, by a pseudo-stereographic projection (Blavette et al., 1982; Cerezo et al., 1999; De Geuser & Gault, 2017).

The first design of atom probes involved using FIM to target regions of interest at the specimen's surface and specifically allow certain imaged atoms to pass through a probe hole in the FIM screen to reveal their elemental identity by time-of-flight mass spectrometry. Already then it became evident that there were so-called 'aiming errors' (Krishnaswamy et al., 1975) that meant that atoms imaged by FIM would not end up being analysed as they 'missed' the hole. The development of the imaging atom probe (Panitz, 1973), with a field of view comparable to FIM but with a limited analytical range, demonstrated that the imaged position of a surface atom by FIM and the impact position of this same atom following field evaporation were dissimilar.

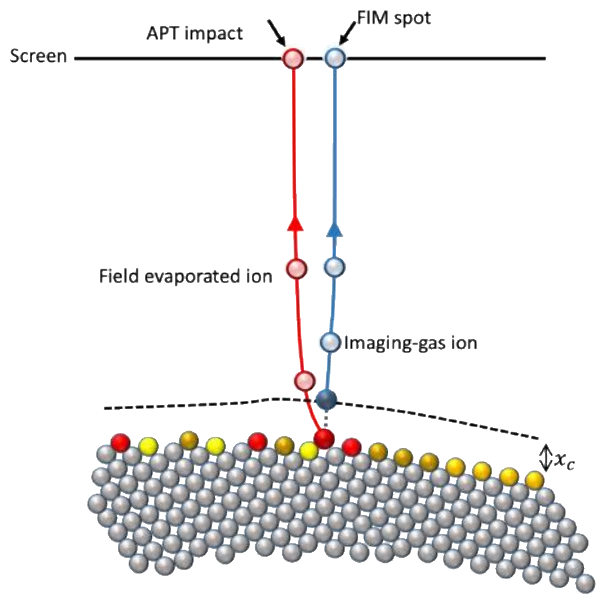

Figure 1: schematic diagram of the difference between the trajectories of a population of imaging gas ions, in blue, leading to the formation of an individual spot on a field-ion micrograph, and, in red, for an individual field evaporated ion leading to a single detector impact in atom probe tomography.

These differences can be understood as image gas ions forming the field-ion micrograph originate from a region located 1–5 Angstroms above the surface typically, near the critical distance for ionisation $x_c$, while as the specimen's atoms ionise and desorb, they do so at the surface. The ions generated by field ionisation and field evaporation hence do not travel through the same electrostatic field. This is summarised in Figure 1. It is in the early stages of its flight, departing from the surface, that the ion is most subject to aberrations. Indeed, right at the surface, the atomic roughness leads to very strong local variations of the electrostatic field that are well reproduced by finite-element simulations (Vurpillot, Bostel, & Blavette, 2000). These highly localised gradients can cause very local surface rearrangements by short-range surface migrations (Waugh et al., 1976), and it was proposed that atoms in the process of leaving the surface 'roll-up' on their neighbours (Schmidt et al., 1993), and demonstrated this experimentally for Rh (Suchorski et al., 1996). Sanchez et al. used density-functional theory (DFT) under intense electrostatic fields to study the energy barrier for field evaporation and field-assisted surface diffusion in the case of Al, and showed that they were linked (Sanchez et al., 2004) – i.e. the magnitude of the electric fields necessary for field evaporation also greatly facilitate surface diffusion, facilitating the roll-up. More recently, Ashton et al. further demonstrated by DFT that the roll-up on neighbours was the energetically favourable path to field evaporation in the case of W (Ashton et al., 2020). These studies on pure metals with the evaporation field ranging from low (Al) to medium (Rh) to high (W) suggest that the roll-up may be affecting most metals, which agrees with the suggestion of Wada (Wada, 1984) that above a relatively modest temperature, the DC field evaporation rate is controlled by surface diffusion processes.

Whereas APT's depth resolution is sufficient to resolve interatomic distances across a wide range of families of planes, materials, and experimental conditions (Vurpillot et al., 2001; Baptiste Gault et al., 2010; Cadel et al., 2009; Jenkins et al., 2020), these effects combine to limit the lateral resolution of APT generally to more than an interatomic distance based on simulations (Vurpillot, Bostel, Cadel, et al., 2000; Vurpillot & Oberdorfer, 2015) and experimentally (Haley et al., 2009; Vurpillot et al., 2001).

These studies were mostly theoretical or on pure metals. Yet APT is nowadays used mostly by materials scientists and engineers interested in mapping the composition of small secondary phase particles or localised compositional fluctuations associated with solute clustering or short-range ordering. Assessing the spatial performance of APT in this context is a true challenge. Recently, De Geuser and Gault performed a systematic review of the literature comparing small-angle scattering to APT and proposed that the spatial resolution in the analysis of nearly spherical particles was limited to somewhere in the range of 0.5nm – 1.5 nm (De Geuser & Gault, 2020). Their study highlighted the complexity to pinpoint an individual value as the resolution will depend on the difference in the field evaporation behaviour of the matrix and that of the precipitate. This approach is however not directly transferable to all analyses, in particular localised segregations and compositionally complex solid solutions, i.e. high-entropy alloys.

In the analyses of these alloys, it is common to study the statistical distribution of alloying elements in an APT dataset in order to compare it to a random distribution, for instance by binning the data into blocks of a certain number of atoms (Hetherington et al., 1991; Moody et al., 2008). This approach is often referred to as 'frequency distribution', and it became commonly used to study compositional fluctuations in spinodally-decomposing systems, i.e. over length scales in the range of nanometres. Its applicability to solute clustering or short-range ordering, i.e. over interatomic distances, has not been assessed. Yet, a cornucopia of articles on bulk metallic glasses (Miller et al., 2003; Kontis et al., 2018; Sarker et al., 2018), nanocrystalline alloys (Detor et al., 2005), high-entropy alloys (Deng et al., 2015; Rao et al., 2017) and other compositionally complex alloys, as well as semiconductors (Galtrey et al., 2007) use it to assess an absence of solute clusters or short-range order.

An alternative method is to directly quantify the distance between nearest neighbours (Shariq et al., 2007; Stephenson et al., 2007) and compare the distribution of distances to the one obtained from a randomly labelled dataset. The latter is a separate, duplicated reconstructed dataset in which the atomic positions are maintained, but the mass-to-charge ratios are swapped randomly, mimicking a randomly distributed set of atoms. The comparison of these two distributions is only typically visual and rarely quantitatively assessed statistically.

Here, we want to offer some reflections and new insights into the spatial performance of APT, based on experimental and computational results, in part based on (in)direct comparison with FIM. The renewed interest in FIM (Vurpillot et al., 2017), including combined with new simulation approaches (Klaes et al., 2020; Katnagallu et al., 2018, 2019) (Klaes et al., 2020; Katnagallu et al., 2018, 2019), and the development of the analytical-FIM (Katnagallu et al., 2019) enable us to provide a critical perspective on the analysis of local neighbourhoods by APT in the case of pure metals, local segregations in dilute alloys and concentrated alloys.

## 2. Methods
### 2.1 Ion trajectory modelling

Here two sets of simulations are reported, both make use of the field evaporation simulation framework introduced in ref. (Rolland et al., 2015). This approach is meshless and uses the Robin model to determine the local charges on each atom at the surface of a field emitter, and then derives the electric field distribution at and in the vicinity of the surface. The simulations of the field evaporation behaviour of the NiRe binary alloy and the concentrated alloys were performed using the latest version of this model described in detail in (Klaes et al., 2020).

In addition, recently, Klaes et al. included the possibility to perform field-ion micrograph simulations (Klaes et al., 2020). This model allows for simulating the trajectories of both the image gas ions departing from the ionisation zone above the specimen's surface and the field evaporated ions departing from the surface itself. The details of the simulation technique can be found in (Klaes et al., 2020). The set of simulations reported for FIM make use of this model.

### 2.2 Analytical FIM

In (Katnagallu et al., 2019), a new approach termed analytical field-ion microscopy (aFIM) was introduced that makes use of the single-particle detector of a commercial atom probe (the LEAP 5000 XS) and the associated time-of-flight mass spectrometer to perform field-ion microscopy under relatively low imaging gas pressure. Here, pure tungsten needles were prepared from a drawn wire most often exhibiting a <110> z-axis orientation, by using electrochemical polishing at 5-8 VAC in a 5% molar NaOH solution. Pure He was introduced in the LEAP analysis chamber, following flushing and evacuating the gas mixing chamber twice with pure gas, and refilling it to approximately 1 Torr (with 1 torr approx. 133 Pa). For FIM, the specimen temperature was set to 50K and, with a manual leak valve, a low pressure in the range of $1 \times 10^{-7}$ Torr of He then admitted into the analysis chamber. The LEAP was operated in high-voltage (HV) pulsing mode at a pulse repetition rate of 250kHz. Field ionised events are recorded on the particle detector and the detector coordinates are accessible through the EPOS file format. Matlab and Python scripts were then used to process the data, scripts which would be made available upon request to the authors.

## 3. Results
### 3.1 Pure metals: experiments

Analytical FIM combines the imaging capability of FIM with the possibility of detecting the ion following its field evaporation, thereby estimating the difference in the imaging ions and corresponding field evaporated atom directly accessible. An aFIM analysis of a pure W specimen is shown in Figure 2. First, an equivalent FIM image is built by forming a histogram of the imaging gas ion detector hits Figure 2a. The typical ring features appear around the main sets of crystallographic planes, as expected. We plot in Figure 2b the location of the $W^{3+}$ and $W^{4+}$ ions that appeared as part of the multiple hits. As expected from the early work of (Waugh et al., 1976) comparing FIM and field desorption, the atomic sharpness of the image is lost. The pattern formed is close to the typical desorption pattern observed in APT for higher charges states and multiple hits detected in the analysis of pure W.

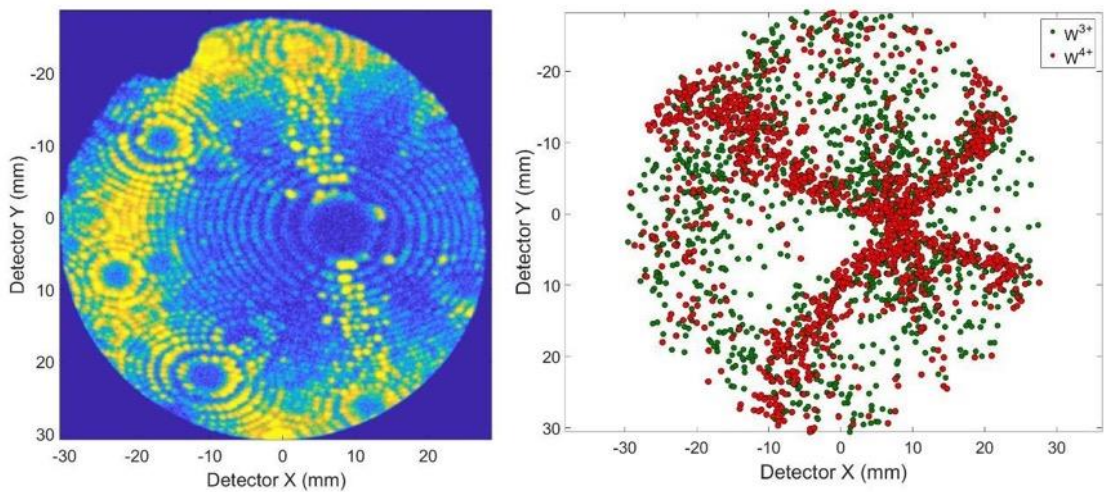

*Figure 2: Image formed from field ionised events on the detector (left) and hit positions of the identified W ions on the detector (right) consistent with field desorption maps seen from APT runs of the higher charge states for pure W.*

We acquired another aFIM dataset on pure W also at 50K on the LEAP 5000 XS, with a 35% pulse fraction over a standing 6kV, and 0.15 to 3.6 x $10^{-8}$ Torr of He. Filtering techniques for the time-of-flight spectrum in aFIM are described in (Katnagallu et al., 2019) and enabled by the correlations in the field evaporation process: the local rearrangements of the charges following the field-induced desorption of an imaging gas atom adsorbed on the surface causes a sudden increase in the local electrostatic field that the neighbouring surface atoms are subjected to (Katnagallu et al., 2018), which favours their field evaporation in close spatial and temporal correlation (De Geuser et al., 2007).

Figure 3a is the corresponding mass spectrum from this dataset, with all the data shown in black, and the data following filtering specifically for ions on multiple hits within selected ranges of mass-to-charge ratio and for impacts that are within 4mm of each other on the detector. The vast majority of multiple hits processed are Ne-W or He-W (>92.9%), and not just W-W multiple hits. This trend is not surprising considering the higher gas pressure. It is also documented that the probability of multi-hit detection is directly proportional to the strength of the electrostatic field and hence the detection rate (De Geuser et al., 2007). Here, the charge state ratio of W (taken as the number of $W^{3+}$ ions detected over $W^{4+}$) in the case of APT (15% PF, 50K) is 30.44 while in the aFIM case (35% PF, 50K, $10^{-8}$ torr of He) it is only 9.52. This indicates that we are under 5–10% lower electrostatic field conditions according to Kingham's calculations (Kingham, 1982). This was to be expected from Müller's early work on the influence of gas on the field evaporation (Müller et al., 1965).

It is apparent from the filtered mass spectrum that some Ne remained in the FIM gas mixing chamber. Ne has a lower ionisation field than He, and cannot be used to image high evaporation field materials like W, as the ionisation occurs too far from the specimen's surface to lead to a high-resolution image (Nishikawa & Muller, 1964). Ne can however adsorb on the cold specimen's surface and migrate up the shank towards the highest electric field regions. Adsorbates on top of a W atom is expected to strongly attract and localise the charges (Neugebauer & Scheffler, 1993), which can have two consequences. First, upon field desorption of the Ne, the redistribution of these charges will cause the neighbouring W atoms underneath to be subject to a higher electrostatic field, thereby enhancing their probability of field evaporation (Katnagallu et al., 2018). Second, both the He or Ne adsorbate and the W atom depart together, possibly aligned along the field line to maximize polarization, and fall apart early during the flight, on the way to the detector, once they have acquired enough charge to be subject to Coulomb explosion (Tsong, 1985; Blum et al., 2016; Zanuttini et al., 2017; Peng et al., 2019). It is worth nothing though that the difference in mass by a factor of 9 and 20 with Ne and He respectively, makes the trajectory of the W least likely affected by Coulomb repulsion associated with a dissociative event. Either of these mechanisms would result in strong multi-hit correlations (De Geuser et al., 2007; Saxey, 2011; Yao et al., 2013), i.e. the detection of a He/Ne ion is often associated with the detection of a W ion from very close locations on the detector.

In Figure 3b, we plotted a detector hit density map using approx. 280 million ion impacts, similar to Figure 2a. Albeit with a slightly coarser binning and with a lower gas pressure that affects the imaging conditions, poles are clearly visible. We superimposed onto this map a quiver plot visualising the average distance between the impact of a $W^{3+}$ ion and a $Ne^+$ generated by the same HV pulse over the entire analysis. The corresponding vector is typically oriented radially with respect to the centre of the nearest large terrace, pointing inward, i.e. the $W^{3+}$ impact is farther from the terrace's centre. Figure 3c maps the standard deviation of the displacement between hits that can be up to 0.4 mm. The distance between impacts is low near sets of atomic planes with high-Miller indices, i.e. on which the atomic packing is relatively loose and the image resolution in FIM is sufficient to distinguish individual atoms (Chen & Seidman, 1971). The longer distances appear near to the (011) set of planes, i.e. the denser planes with the wider spacing and hence the widest terraces that lead to the build-up of stronger electrostatic gradients.

A quantitative analysis of these aberrations is out of the scope of the present article, in part because to be meaningful, this would require a thorough and systematic investigation, but also because the actual magnitude of the aberrations will vary over the course of the analysis as the specimen shape and the magnification evolve and is subject to substantial variations across the field of view as highlighted in Figure 3–c. It must also be pointed out that although the charge-state ratio indicates electrostatic fields in the same range, opposite to when laser pulsing is used for instance (Gault, Chen, et al., 2011), the presence of the gas can modify the field evaporation conditions, and to quote Erwin Müller: "The gas-surface interactions at the emitter tip of a field ion microscope are quite complicated, and a quantitative understanding is difficult because of the many uncertainty factors of thermal accommodation, particularly in the presence of an adsorbed layer" (Müller et al., 1965).

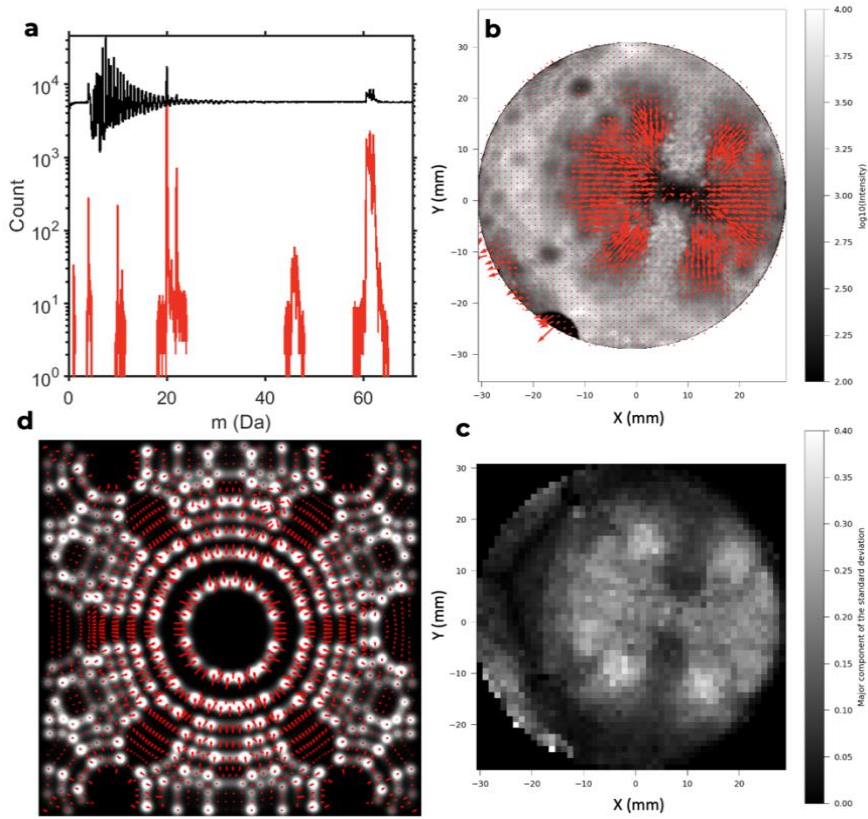

*Figure 3: (a) unfiltered (black) and filtered mass spectra for an aFIM experiment from pure W in He at 50K. (b) Quiver plot of the distance from the impact of a $W^{3+}$ atom to a $Ne^+$ (approx. 10 million pairs), superimposed on a recalculated equivalent FIM image (280 million hits), and (c) map of the standard deviation of the associated displacement. (d) FIM image simulation in black & white, and superimposed quiver plot with vectors starting from an ion impact position and pointing to its corresponding location on the FIM image.*

### 3.2 Pure metals: computational study

As expected from modelling work, the aberrations are intimately linked to the atomic-scale structuring of the specimen's surface, i.e. terracing and local neighbourhoods. In order to directly compare our experimental observations with simulations, we generated needle-shaped synthetic data for a pure metal with a body-centred-crystal structure, i.e. similar to W. The specimen's radius was 12 nm. This was used as input for simulating FIM images and APT data.

Figure 3d shows part of a simulated FIM image centred on the (011) terrace in white, with individual atomic positions imaged. A quiver plot is superimposed, with red vectors starting from the location of the impact of the ion following its field evaporation and propagation from the specimen's surface onto the virtual detector and pointing to the corresponding position of the imaged atom in the FIM simulation. A similar set of observations can be made: aberrations are mostly centrifugal with respect to the terraces, and low index poles lead to more pronounced aberrations. A key message from these results is that the regions primarily badly affected by trajectory aberrations are the low-density areas on the desorption image, which are typically where the atomic planes can be imaged in the depth of the reconstructed dataset in APT (Moody et al., 2009) and FIM (Vurpillot et al., 2007; Klaes et al., 2020). Higher-index poles are however expected to exhibit a lower depth resolution (Gault et al., 2009). These simulations point to a necessary compromise between the lateral and depth resolution.

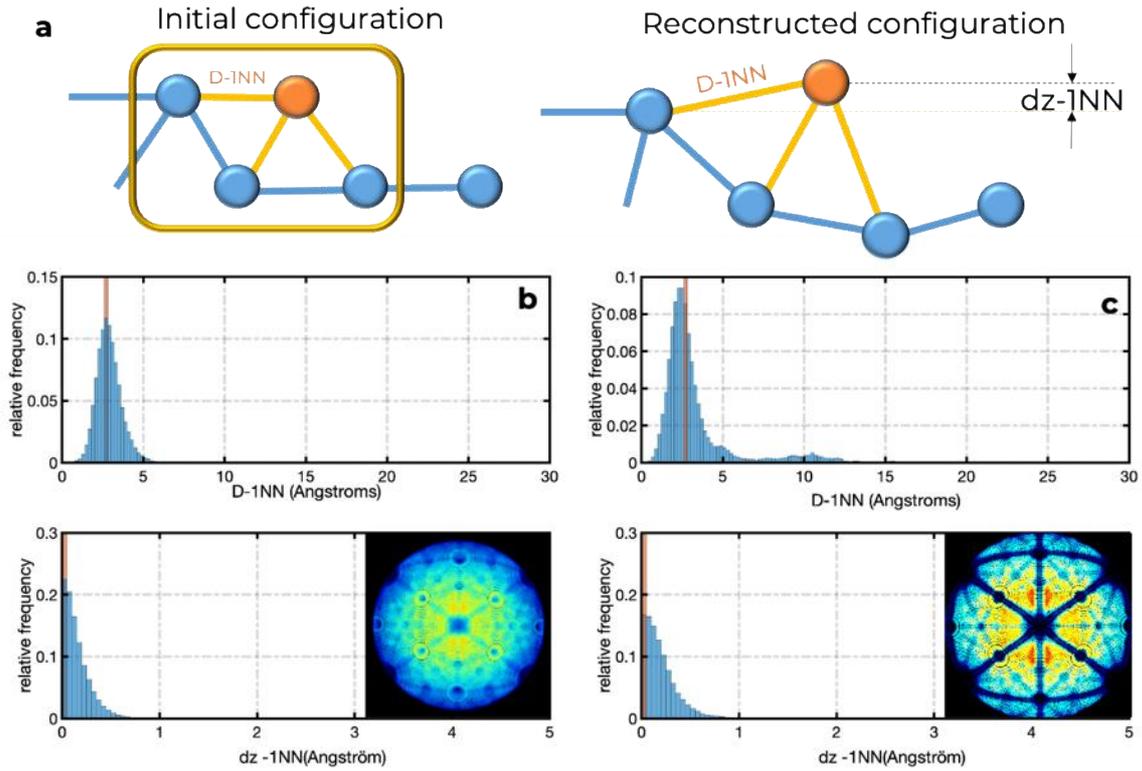

*Figure 4: (a) schematic of the initial and reconstructed atomic configurations, and the calculated quantities: the first near-neighbour distance D-1NN and the z-offset dz-1NN. D1NN and dz-1NN for atoms initially on the same plane in the input data in (b) a reconstructed FIM dataset (inset is the multilayer detector map); and (c) the corresponding reconstructed APT simulated data. Note that the orange bar in each histogram represents the initial D-1NN and dz-1NN distributions from the same volume of a perfect lattice, whereas the blue histograms represent the measured D-1NN and dz-1NN histograms of distances (using the indexes of the 1st nearest neighbours) in the reconstructed volumes.*

In an effort to quantify the influence of these aberrations on the spatial performance of APT, we performed a series of calculations schematically depicted in Figure 4a. We consider the distance to all atoms in the first shell of nearest-neighbour atoms sitting on the same plane, which we refer to as D-1NN. These distances are extracted from a Delaunay tessellation (Lefebvre et al., 2011; Felfer et al., 2015) and calculated for all ions in a reconstructed dataset that covers an angular field-of-view of +/- 30° around the [011] direction in the centre of the virtual detector. The D-1NN in the input simulation cell can be directly compared to the distance to this very same nearest-neighbour in the reconstructed data. We also extract more specifically the offset in depth, dz-1NN. This offset is related to the angular difference in their trajectory, but also to the sequence in which the ions are detected, i.e., when two atoms evaporate rapidly after one another, the dz is typically low – see (Vurpillot et al., 2013) for more details. Here, we perform this calculation to estimate the spatial dispersion induced by the imaging process for a FIM image simulation for a body-centred cubic crystal, and for atoms on the same (011) atomic plane and for a field-of-view similar to experimental data.

The histograms of distances to the first nearest-neighbour and z-offset are plotted in Figure 4b. In orange is the distribution in the input simulation cell, prior to the imaging and evaporation simulation. All distances are in a single bin, which reflects the undistorted value in the original bcc crystal. In blue is the distribution in the 3D FIM reconstructed volume using the protocol introduced by (Klaes et al., 2020). The distribution is rather narrow around the expected theoretical value, with a full-width half maximum (FWHM) of approx. 0.15nm, i.e.

around half of the distance to the first nearest neighbours, making it possible to separate the two neighbouring atoms, i.e. all atoms are within twice the initial nearest-neighbour distance. The dz-1NN also shows a narrow spread of approx. 0.015 nm FWHM, emphasising once again the better spatial resolution in depth compared to laterally (Klaes et al., 2020). In the inset is the map representing the pile-up of positions of FIM atomic spot centres obtained after the evaporation of several surface layers. The colour scale corresponds to the ion impact density, calculated around each impact using the number of counts in a delimited circle, with a radius equivalent to about 1 nm at the tip surface.

Figure 4c shows the same histograms obtained from the corresponding APT simulation, inset is the detector impact map. The picture here is very different. The FWHM may not have changed much, still near 0.15nm. However, the peak position has shifted towards a lower distance (0.2nm instead of 0.22nm, i.e. approx. 10% difference) from the theoretical value, which indicates significant density fluctuations (Stephenson et al., 2007). More worryingly, nearly 17% of the ions have landed at a distance twice or more than the expected value, making it very unlikely that two neighbours from within the initial volume are indeed neighbours in the reconstructed data. The spread in distances is further evidenced in the graphs shown in log scale (see Suppl. Information).

### 3.3 In Alloys: simulations

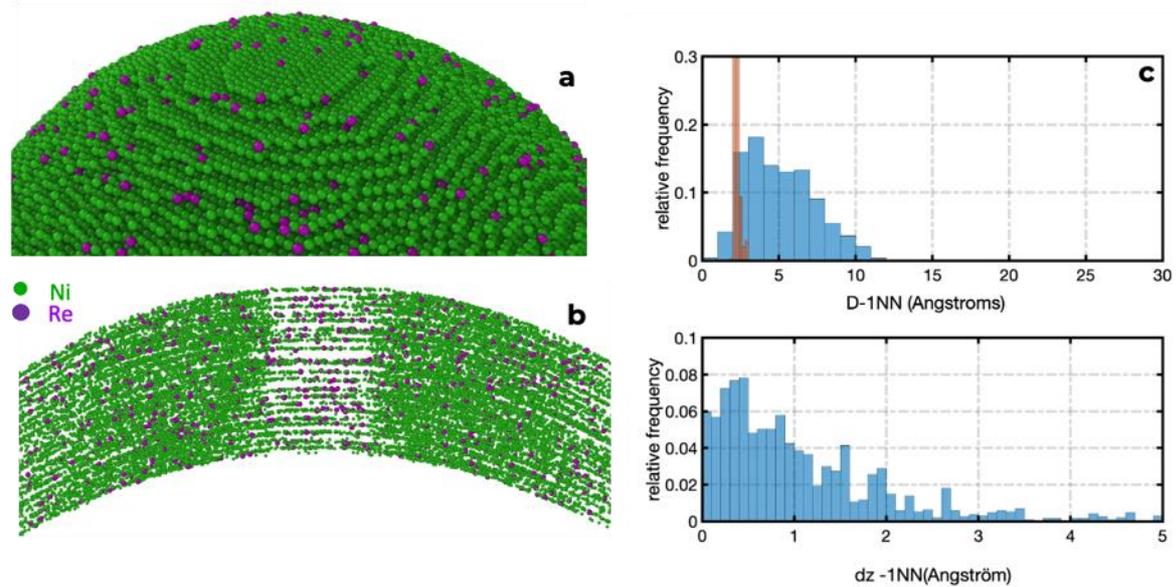

*Figure 5: (a) simulation cell with Re segregated to stacking faults in Ni and (b) thin slice though the reconstructed simulated data. (c) D1NN and dz-1NN for Re atoms initially on the same plane in the input data in orange and in the simulated and reconstructed data in blue.*

In order to assess these effects in the case of alloys, we revisited the APT simulations performed in the study of a Ni-2% Re alloy by APT and analytical-FIM and reported in (Katnagallu et al., 2019). In this case, the Re atoms are simulated by a high-field solute segregated to stacking faults formed by deforming pure Ni by molecular dynamics in the large-scale atomic/molecular massively parallel simulator (LAMMPS). 20% of the sites on the stacking faults are replaced by atoms with a 30% higher evaporation field than the matrix. A view of the simulation cell before field evaporation is shown in Figure 5(a), with the Ni matrix atoms in green and the high-field Re atoms in purple. Robin-Rolland simulation code was used on a 12 nm radius NiRe modelled tip using basic single evaporation field constants for Ni and Re, at T=0K. Impacts on a detector placed at 1 micron in front of the sample were used to reconstruct a virtual APT volume. As reported in (Katnagallu et al., 2019), reconstruction parameters were optimized to have a volume with the correct initial known dimensions. Figure 5(b) is a thin slice through the data reconstructed following the simulation of the field evaporation process. The atomic planes are imaged parallel to the tangent to the local reconstructed curved surface across the field of view, i.e. (002) planes in the centre and the (022) planes towards the edges of the volume.

Figure 5(c) show the histograms of the nearest-neighbour distance and z-offset, respectively, between a Re atom and its first nearest neighbour Re atom. In this case, almost half of the Re atoms that were initially first neighbours end up shifted by more than twice the first nearest neighbour distance. The distribution in the input data, in orange, is not infinitely narrow because of the defects that shift atoms from their ideal positions. The difference with the distribution of the nearest-neighbour distance in the reconstructed data in blue is readily visible: some atoms are reconstructed over a nanometre away from each other, 5 or more interatomic distances away from where they were initially.

With regards to the z-offset specifically, and conversely to the distributions in Figure 4c, the distribution is substantially distorted. This can be ascribed to a combination of effects. First, preferential retention of the Re with a higher evaporation field than the Ni matrix modifies the sequence of detection and causes Re atoms to be reconstructed deeper than they should have been. Second, the assumed curvature of the emitter in the reconstruction protocol means that two initially neighbouring atoms will be reconstructed at increasingly different depths as their respective impact positions are farther from each other (Gault, Haley, et al., 2011).

### 3.4 Concentrated solid solutions

Here, we wanted to simulate the case of a high-entropy alloy, which can also be referred to as a compositionally complex alloy or a concentrated multi-component solid solution. These alloys are the focus of many studies at the moment, and one of the most widely studied compositions is an equiatomic mixture of Fe, Cr, Ni, Co and Mn. These elements are expected to have close evaporation fields in their pure form, nearly all within 25% around 30V.nm$^{-1}$(Tsong, 1978). The evaporation field is species dependent but also depends on the local neighbourhood at the specimen's surface (Ge et al., 1999). An approach to model this is to assume a single average evaporation field, modulated randomly to mimic the different species. As input for the simulations, we hence assumed an equiatomic mixture of atoms from five species, randomly distributed on a body-centred cubic lattice. Each species is given a specific evaporation field in the range +/- 5% , +/- 10% and +/- 20% around an average value. A face-centred cubic lattice would have been closer to most studied alloys, but based on similarities in the aberration patterns from simulations from the different crystal structures (Oberdorfer et al., 2015), we expect that qualitatively similar results would have been obtained. Robin-Rolland simulation code was used on a 25 nm radius modelled tip at T=0K.

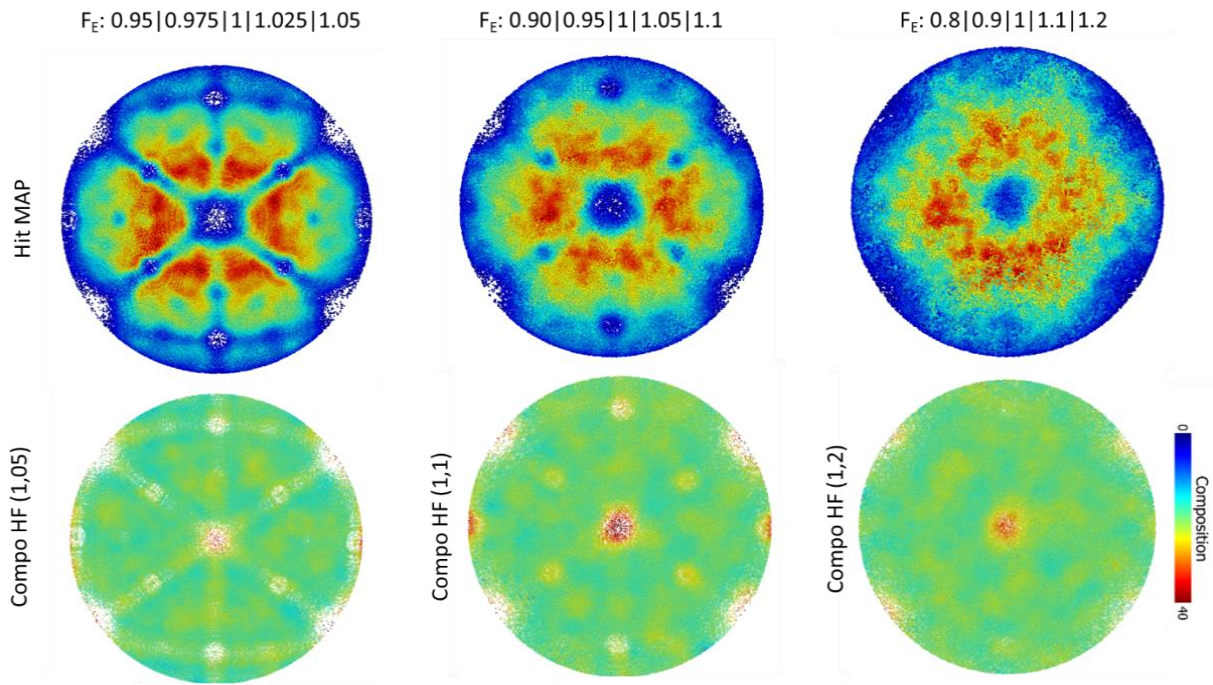

*Figure 6: overall density map and composition map for the element of highest evaporation field for simulated data with elements with evaporation fields in the range (a) 0.95–1.05, (b) 0.9–1.1, (c) 0.8–1.2.*

Figure 6 summarises the results from the simulations for the +/- 5%, +/- 10% and +/- 20% ranges, respectively in (a–c). The average composition is close to the expected 20% for all elements except in specific regions near poles, where the elements of higher evaporation field are much more highly concentrated – up to approx. 40%. Similar issues had been reported in the past in the analysis of Al-alloys for instance (Baptiste Gault, Moody, Cairney, & Ringer, 2012), and it was often thought to be related to surface migration (B. Gault et al., 2012), which are not accounted for in our model, conversely to others (Gruber et al., 2011), and hence cannot explain these results. This is related to the specific retention on the surface of the field evaporating specimen of the species of highest evaporation field until the local curvature near atoms of that species allows for reaching a sufficient electric field to cause its departure. Since the sequence in which ions are detected is used to derive the z-coordinate of each atom during the data reconstruction process, such a retention effect will not only affect the apparent composition but also the depth resolution.

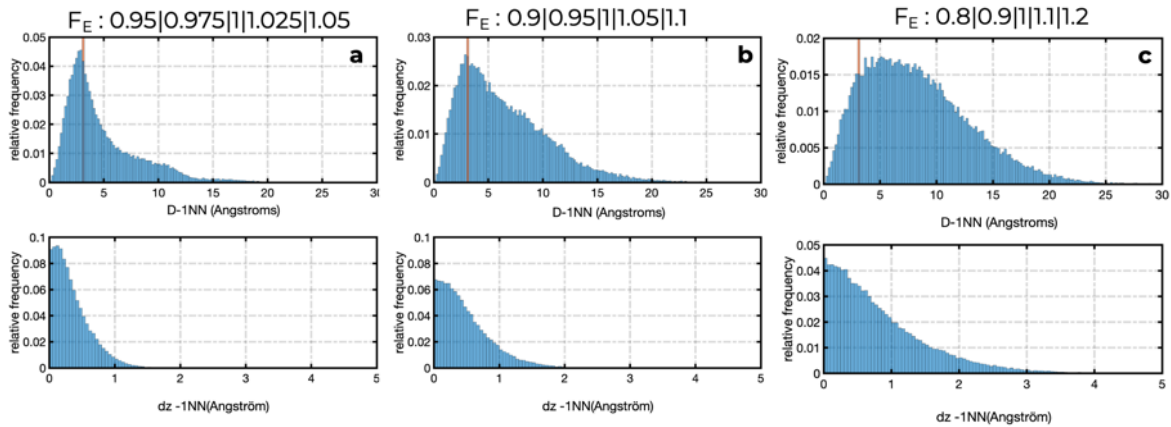

*Figure 7: D-1NN and dz-1NN distributions between atoms initially nearest neighbours and located at the same depth in the input data after field evaporation simulation and data reconstruction for a range of evaporation field of (a) +/- 5%; (b) +/- 10%; and (c) +/- 20% around an average field.*

Figure 7 reports the distance and z-offset distributions between atoms initially nearest neighbours in the input data for the three ranges of evaporation fields +/- 5% , +/- 10% and +/- 20%. The FWHM of the D-1NN distribution changes substantially, increasing from approx. 0.3nm to 0.75 nm to 1.2 nm, and in depth (z-offset) from 0.04nm to 0.075 nm up to 0.1 nm as the range of evaporation field increases. The precision of the measurement hence worsens with the compositional complexity increasing, with a clear mixing of nearest-neighbour positions. The depth coordinate is affected but not as much and the depth resolution is hence more robust against the distribution of evaporation fields within the material, agreeing with experimental observations.

## 4 General discussion

Across the different cases we have studied here, we revisited two important aspects of APT analysis: (i) the spatial resolution is not isotropic, and the depth resolution remains higher than the lateral resolution; and (2) the depth resolution is optimal near the poles, that is in very specific locations within a dataset and not across the entire field-of-view. The depth resolution mostly relates to how sequential the field evaporation proceeds, which can be ascribed to how well the field localizes on the edges of terraces. On dense planes, with a wide interplanar spacing, terraces are wide and the unravelling of the atomic structure by field evaporation follows a well-defined sequence. On open planes, the concentration of the field is less on the edges of the terraces, but on individual atoms. This is what enables true atomic resolution in FIM, but also makes the sequence of evaporation less well determined. This underpins the compromise between lateral and depth resolution that was mentioned above. Evaporation field variations make this balance more difficult to achieve, because they firstly hinder a well-sequenced evaporation, and the atoms with a higher evaporation field remaining on the surface can further cause lateral aberrations. Ultimately, neighbourhood analyses should typically be restricted to where atomic planes can be imaged as they are a sign that the field evaporation process is sufficiently well ordered to maintain the neighbourhood relationships. However, they can also be locations of aberrant compositional analyses. Let us discuss specific aspects.

## 4.1 Pure metals

The results we presented herein are not meant to be dismissive of previously published reports in the literature, but merely to get the reflection going on what information APT can confidently provide. Some of the authors themselves have reported on the study of atomic neighbourhoods (Stephenson et al., 2007) and even on exploiting this information to try and reconstruct lattice positions in pure metals (Vurpillot et al., 2003; Moody et al., 2011; Breen et al., 2015) and alloys (Moody et al., 2014; Gault et al., 2017). A critical perspective should have arisen from the low signal-to-background ratio in the x-y-spatial distribution maps from which the lateral resolution of APT was estimated: even in pure metals, with a narrow distribution of evaporation fields (Yao et al., 2015), only a small fraction of neighbouring atoms is reconstructed at distances compatible with their actual first shell of nearest neighbours. This was evident in (Moody et al., 2009) and was discussed at length also in other reports (Haley et al., 2009).

Instead, we maintained a misleading impression that despite its apparent limitations, the lateral resolution of APT remained sufficient to resolve the lateral extent of atomic arrangements. Many have heard of projects hoping to achieve full crystallographic analysis from APT data, for instance, (Ziletti et al., 2018), especially motivated by the advent of machine-learning and its application in related fields (Butler et al., 2018; Aguiar et al., 2019). However, the physics governing the early stages of the departure of the charged particle from the surface critically limit the lateral resolution of APT. The probabilistic nature of the process makes it extremely challenging, if possible at all, to correct for the resulting aberrations.

## 4.2 Alloys

First, the set of simulations included herein also shows that detection of high-field species at the centre of the poles is not only related to surface diffusion but to the difference in the evaporation field behaviour between high field and low field. This affects e.g. AlCu or AlSi and can be ascribed to 'chromatic aberrations' as discussed in (Marquis & Vurpillot, 2008).

Second, our results in the analysis of solutes segregated in a binary alloy point to limitations both in the lateral and depth coordinates. A 30% difference in evaporation field between solutes and solvent is reasonable. Based on tabulated values of the evaporation field, which is admittedly an approximation, this would be less than the difference between Al and Cu for instance, combinations for which the comparison between small-angle scattering data and APT in the analysis of clusters show strong differences (De Geuser & Gault, 2020). Neighbourhood relationships in these clusters are hence highly unlikely to be maintained and this raises questions as to the pertinence of nearest-neighbour based cluster-search algorithms (Dumitraschkewitz et al., 2018). Over the course of a single experiment and across the field-of-view, the actual distance to a first nearest neighbour can change due to the magnification being non-isotropic and associated with the change in the magnification as the specimen gets blunter due to field evaporation. These aspects make approaches based on radial distribution functions to extract the characteristics of populations of clusters (Zhao et al., 2018) relatively more robust.

## 4.3 Error estimation

We feel it is pressing that the community comes to terms with these limitations and makes use of the technique for its strengths, without attempting to exploit it for what it is not. This should not prevent us from trying to push the performance limits of APT, and challenge its limitations, but keeping in mind that the limits are mostly bound by the physics and not by the way we extract or process the data. For instance, and despite its inherent limitation, there is still much that APT can do, including in the analysis of crystallographic features patterns (Baptiste Gault, Moody, Cairney, Ringer, et al., 2012), by focusing on the analysis of the desorption patterns to work out orientation (Wei et al., 2018) or by using the more highly-resolved information in the depth to investigate site occupancy (Li et al., 2021). The difference in the evaporation field between species in a mixture may also preclude some of these analyses.

The depth-resolution is relatively robust against changes in the base temperature in high-voltage pulsing mode and peak temperature in laser pulsing mode for pure metals (Baptiste Gault et al., 2010; Gault, Chen, et al., 2011; B. Gault et al., 2010). However, the relative difference in the evaporation field between species can differ as a function of the temperature (Wada, 1984), and, hence, the situation could worsen more substantially in specific cases and lead to a critical loss of resolution that was for instance observed in some ordered phases (Vurpillot, Bostel, Cadel, et al., 2000; Boll et al., 2007). Performing these analyses might require careful experimental design, for instance, by preparing specimens along specific orientations as discussed in (Jenkins et al., 2020), optimising the experimental conditions (pulsing mode, base temperature, detection rate), and targeting the search for short-range ordering in metallic alloys along the direction where the depth resolution is potentially sufficient to reveal the signal – i.e. typically low-index sets of planes.

It is also critical that we accept that the precision of our measurement is limited and we should probably get better, as a community, at assessing the spatial error on the atomic positions. For instance, it is commonly accepted that error bars are included in the composition measurement in a composition profile. Even if these errors might be underestimated as they do not account for species-specific losses associated with detector pile-up (Meisenkothen et al., 2015; Peng et al., 2018) or molecular dissociations leading to neutral fragments for instance (Zanuttini et al., 2017; Gault et al., 2016). However, using error bars on the distance axis is most uncommon.

It may be that the representation as a point cloud is misleading – the location at which we reposition a single atom is just one of the possible positions where this atom may have been reconstructed, and it might not be the most probable position where it was located. We do not offer ready-made solutions here, simply point to some of these issues and highlight their complexity, in the hope to raise consciousness and motivate, maybe, studies in this exciting direction.

# 5 Conclusions

To conclude, lateral neighbourhood relationships within materials are modified and typically not maintained by the field evaporation and APT data reconstruction process. Measured relationships should not be readily interpreted, and only directional neighbourhood analyses in the depth may contain relevant information. This was hinted at in previous works, and we provide further evidence here that this affects not only pure metals or metallic glasses, but also when atoms of different evaporation fields are segregated within a matrix, and in compositionally complex alloys. The resolution is more robust in depth, and it is necessary to develop approaches to probe neighbourhoods selectively in this direction. This may enable to exploit the higher resolution in this direction to reveal neighbourhood relationship, e.g.

short-range order. Since the aberrations arise in the early stages of the ionic flight, imaging of the atomic neighbourhoods prior to the field evaporation, by using field-ion imaging and aFIM in particular, bears a lot of potential for the future.

# 6   Acknowledgements

We thank Uwe Tezins, Christian Broß, and Andreas Sturm for their support to the FIB and APT facilities at MPIE. Shyam Katnagallu (now at KIT) and Isabelle Mouton (now at CEA) are thanked for their help and support with the aFIM. L.T.S. and B.G. acknowledge financial support from the ERC-CoG-SHINE-771602. Y.L. and L.T.S acknowledge support from the Max Planck research network on big-data-driven materials science (BiGmax). F.F.M. acknowledges financial support from the International Max Planck Research School for Interface Controlled Materials for Energy Conversion (IMPRS-SurMat). The work was funded through EMC3 Labex BREAKINGAP and the EQUIPEX ANR-11-EQPX-0020 (GENESIS). Simulated samples were generated using the ATOMSK software (Hirel, 2015). Visualization was performed using OVITO software (Stukowski, 2010). FV thanks also the financial support of the University of Rouen through a CRCT funding.